\documentclass[prb,aps,twocolumn,superscriptaddress]{revtex4}
\usepackage{graphicx,amsmath,bm,xcolor}
\usepackage{dcolumn}

\begin{document}

\title{Hidden Bose-Einstein Singularities in Correlated Electron Systems}

\author{Takafumi Kita}
\affiliation{Department of Physics, Hokkaido University, Sapporo 060-0810, Japan}
\date{\today}

\begin{abstract}
Hidden singularities in correlated electron systems, which are
caused by pair fluctuations of electron-electron or electron-hole bubbles obeying Bose-Einstein statistics, are clarified theoretically.
The correlation function of each pair fluctuation is shown to have a bound in the zero Matsubara frequency branch,
similarly to the chemical potential of ideal Bose gases. 
Once the bound is reached, the self-energy starts to acquire a component proportional to Green's function itself,
i.e., the structure called one-particle reducible, to keep the correlation function within the bound.
The singularities are closely related to, but distinct from, phase transitions with broken symmetries.
Passing down through them is necessarily 
accompanied by 
a change in the single-particle density of states around the excitation threshold, such as the pseudogap behavior
found here for the negative-$U$ Hubbard model above the superconducting transition temperature.
\end{abstract}


\maketitle

\section{Introduction}

Systems of identical particles with integer spins undergo a unique phase transition called Bose-Einstein condensation 
characterized by the emergence of
 macroscopic quantum coherence \cite{Cornell95,Ketterle97,Leggett06}, 
 which may be regarded as the realization of the singularity of the Bose distribution function.
Specifically for noninteracting systems below the critical temperature, 
the chemical potential is pinned to the value of the lowest energy level with a singularity,
at which particles accumulate macroscopically.
The purposes of the present paper are to show that similar singularities are also present
in the pair-fluctuation channels of fermion systems described in terms of the Bose distribution function
and to clarify some basic features of the hidden phases realized after the singularities have been reached.

The singularities are relevant to logarithmic contributions to the grand thermodynamic potential given diagrammatically 
in terms of closed electron-electron or electron-hole bubbles (see Fig.~\ref{fig1} below).
The argument of the logarithm can be expressed in terms of boson Matsubara frequencies, 
and the vanishing of it in the zero-frequency branch has been identified as the second-order transition point
in the perturbative treatments of using the bare Green's function, 
known as the Thouless criterion for superconductivity \cite{Thouless60,NSR85} 
and the Stoner criterion for ferromagnetism  \cite{Stoner38,Blundell01}, for example. 
they are also identical with the mean-field estimations of critical temperatures
as well as those based on random-phase approximations \cite{IKK63,BS66,DE66}.
When making Green's function self-consistent and incorporating terms other than the logarithmic contributions, however, 
each singular point generally becomes different from the corresponding transition point with a broken symmetry,
as shown below both analytically and numerically.
It also has not been clarified what happens after the singular points are reached and how to describe the phases theoretically.
These are the issues to be studied here.

This paper is organized as follows. In Sect.\ 2, we elaborate the hidden singularities and 
introduce a procedure to handle them theoretically.
In Sect.\ 3, we present numerical results for the negative-$U$ Hubbard model.
In Sect.\ 4, we provide concluding remarks. We use the units of $\hbar=k_{\rm B}=1$.

\begin{figure}[b]
\includegraphics[width=0.9\linewidth]{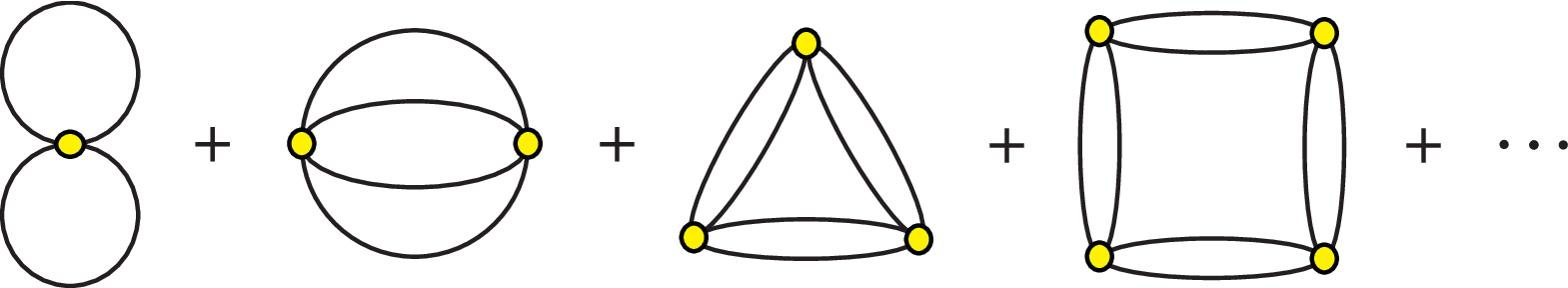}
\caption{\label{fig1}Diagrammatic expression of $\Phi$ in the FLEX-S approximation \cite{Kita11}.
A small circle denotes the interaction vertex with particle-hole and spin degrees of freedom, and a line denotes an element of $\hat{G}$.
}
\end{figure}

\section{Hidden Singularities}

\subsection{Luttinger-Ward functional}

One of the theoretical frameworks suitable for the present purposes is the Luttinger-Ward expression of the grand thermodynamic potential $\Omega$
given as a functional of Green's function $\hat{G}$ \cite{LW60,DDM64,DDM64-2,HRCZ07,Kita11}.
The stationarity condition of $\Omega$ in terms of $\hat{G}$ yields a self-consistent equation for  $\hat{G}$, which has been known, for example, as the Dyson, the Schwinger-Dyson, or Dyson-Gor'kov equation.
Being nonlinear in $\hat{G}$, it can describe ordered phases with spontaneous broken symmetries continuously
from the normal state on the same footing.
Indeed, the ordered phases are characterized by the emergence of 
some novel component in the matrix $\hat{G}$ with particle-hole and spin degrees of freedom,
such as the anomalous Green's function for superconductivity and spin polarization in $\hat{G}$ for ferromagnetism. 
This Luttinger-Ward functional with full particle-hole and spin degrees of freedom can be written as \cite{Kita11}
\begin{align}
\Omega  = -\frac{1}{2\beta} {\rm Tr} \, \bigl[\ln \bigl(-\hat{G}_0^{-1}+\hat\Sigma\bigr) +\hat\Sigma \hat{G} \bigr]+\Phi,
\label{Omega}
\end{align}
where $\beta$ is the inverse temperature, 
$\hat{G}_0$ is the noninteracting Green's function,
$\hat\Sigma$ is the self-energy, and Tr denotes the trace over the arguments of $\hat{G}$.
The stationarity condition of Eq.\ (\ref{Omega}) with respect to $\hat{G}$ yields the Schwinger-Dyson equation:
\begin{align}
\hat{G}=\bigl(\hat{G}_0^{-1}-\hat\Sigma\bigr)^{-1}\hspace{5mm}\mbox{with}\hspace{5mm}\hat\Sigma=2\beta\frac{\delta\Phi}{\delta \hat{G}} .
\label{SD}
\end{align}
These are formally the exact expressions, and practical calculations can be performed on the basis of some approximate expression
for $\Phi\!=\!\Phi[\hat{G}]$, obeying Eq.\ (\ref{SD}) similarly to the exact theory.
The scheme of Eq.\ (\ref{SD}) has also been known as conserving approximations of Kadanoff and Baym \cite{KB62} or $\Phi$-derivable approximations
 \cite{Baym62}. 
The potential $\Omega$ as a functional of $\hat{G}$ is alternatively called the {\it effective action}
 in relativistic quantum field theory \cite{CJT74,Weinberg95}.

\subsection{Normal state}
 
To elucidate the hidden singularities that lie near phase transitions with broken symmetries,
we specifically consider the negative-$U$ Hubbard model at a low electron density $\bar{n}\!=\! N/V$,
which has negligible lattice anisotropy and definitely exhibit $s$-wave superconductivity at low temperatures.
We study this system on the basis of the fluctuation-exchange approximation for superconductivity (FLEX-S approximation),
whose $\Phi$ is given diagrammatically in Fig.~\ref{fig1} \cite{Kita11}.
We first focus on the normal state, where the series reduces to the $\Phi$ functional of the standard fluctuation-exchange (FLEX) approximation:\cite{BSW89,BS89,Kita11}
\begin{align}
\Phi_{\rm n}=&\,\frac{UN^2}{4V}+\frac{1}{\beta}\sum_{\vec{q}}\Bigl[\ln(1-x)+x\Bigr]_{x=U\chi_{G\bar{G}}^{}(\vec{q}\,)} 
\notag \\
&\,  +\frac{3}{2\beta}\sum_{\vec{q}}\biggl[\ln(1-x)+x+\frac{1}{2}x^2\biggr]_{x=U\chi_{GG}^{}(\vec{q}\,)}
\notag \\
&\,  +\frac{1}{2\beta}\sum_{\vec{q}}\biggl[\ln(1+x)-x+\frac{1}{2}x^2\biggr]_{x=U\chi_{GG}^{}(\vec{q}\,)}\, .
\label{Phi_n}
\end{align}
Here, $\vec{q}\equiv({\bm q},i\omega_\ell)$ indicates the four-momentum consisting of the wavevector ${\bm q}$ 
and boson Matsubara frequency $\omega_\ell$. The functions $\chi_{GG}^{}(\vec{q}\,)$ and $\chi_{G\bar{G}}^{}(\vec{q}\,)$
are defined in terms of the normal-state Green's function $G(\vec{k})$ and its time-reversed partner
$\bar{G}(\vec{k})\!\equiv\! G(-\vec{k})$
generally by
\begin{subequations}
\label{chi_AB}
\begin{align}
\chi_{AB}^{}(\vec{q}\,)\equiv&\, -\frac{1}{\beta V}\sum_{\vec{k}}A(\vec{k}+\vec{q}\,)B(\vec{k}),
\label{chi_AB-def}
\end{align}
where $\vec{k}\!\equiv\!({\bm k},i\varepsilon_n)$ with $\varepsilon_n$ denoting 
the fermion Matsubara frequency; this function satisfies
\begin{align}
\chi_{AB}^{}(\vec{q}\,)\!=\!\chi_{BA}^{}(-\vec{q}\,)\!=\! \chi_{\bar{A}\bar{B}}^{}(-\vec{q}\,).
\label{chi_AB-symm}
\end{align}
\end{subequations}
as can be shown by a change in the summation variable $\vec{k}$.
The first term in Eq.\ (\ref{Phi_n}) is the Hartree-Fock contribution,
the second one is the particle-particle ladder series relevant to the superconducting fluctuations  including the second-order contribution,
and the third and fourth terms represent the spin and charge fluctuations, respectively,
composed of the particle-hole ladder and ring diagrams \cite{BSW89}.
The corresponding self-energy is obtained by the functional differentiation of Eq.\ (\ref{SD});
for the normal state that is symmetric in the spin and particle-hole spaces, the differentiation is simplified to 
$\Sigma_{\rm n}(\vec{k})\!=\!\frac{\beta}{2}\frac{\delta\Phi_n}{\delta G(\vec{k})}$
in terms of Eq.\ (\ref{Phi_n}), yielding \cite{BSW89,BS89,Kita11,Yanase03,Kontani13}
\begin{align}
\Sigma_{\rm n}(\vec{k})=&\,\frac{UN}{2V}
+\frac{U}{\beta V}\sum_{\vec{q}}\frac{U\chi_{G\bar{G}}^{}(\vec{q}\,)}{1-U\chi_{G\bar{G}}^{}(\vec{q}\,)}G(-\vec{k}+\vec{q}\,)
\notag \\
&\,+\frac{3U}{2\beta V}\sum_{\vec{q}}\frac{\bigl[U\chi_{GG}^{}(\vec{q}\,)\bigr]^2}{1-U\chi_{GG}^{}(\vec{q}\,)}G(\vec{k}-\vec{q}\,)
\notag \\
&\,-\frac{U}{2\beta V}\sum_{\vec{q}}\frac{\bigl[U\chi_{GG}^{}(\vec{q}\,)\bigr]^2}{1+U\chi_{GG}^{}(\vec{q}\,)}G(\vec{k}-\vec{q}\,).
\label{Sigma_n}
\end{align}

Now, we are ready to discuss the hidden singularities in full detail.
With $G$ replaced by $G_0$ in Eq.\ (\ref{Sigma_n}), 
the conditions of the vanishing denominators at $\vec{q}=0$ in the second and third terms
reproduce the Thouless criterion $U\chi_{G_0\bar{G}_0}^{}(\vec{0})\!=\!1$ for superconductivity \cite{NSR85}
and the Stoner criterion $U\chi_{G_0G_0}^{}(\vec{0})\!=\!1$ for ferromagnetism  \cite{Blundell01},  respectively.
However, the renormalized conditions $U\chi_{G\bar{G}}^{}(\vec{0})\!=\!1$ and $U\chi_{GG}^{}(\vec{0})\!=\!1$
do not specify the phase transition points of respective broken symmetries.
Indeed, by linearizing the self-consistent equation for the anomalous Green's function $F$ in the FLEX-S approximation,
we obtain the equation for the superconducting transition temperature $T_{\rm c}$
that the eigenvalue problem
\begin{align}
\sum_{\vec{k}'}\left[\delta_{\vec{k}\vec{k}'}+\frac{1}{\beta V} U^{\rm s}_{\vec{k}\vec{k}'}G(\vec{k}')G(-\vec{k}')\right]\Delta(\vec{k}')=0
\label{Tc-eq}
\end{align}
has the non-trivial anomalous self-energy $\Delta(\vec{k})$ due to the emergence of a zero eigenvalue, where $U^{\rm s}_{\vec{k}\vec{k}'}$ is the pairing interaction with correlations given  explicitly by Eq.\ (91) in Ref.\ \onlinecite{Kita11},
i.e.,
\begin{align}
U_{\vec{k}\vec{k}'}^{\rm s}
=&\,U\!\left[1-2x+\frac{3}{2}\frac{x}{1-x}-\frac{1}{2}\frac{x}{1+x}\right]_{x=U\chi_{GG}^{}(\vec{k}-\vec{k}')}
\notag \\
&\,-\frac{U^{2}}{\beta V}\sum_{\vec{q}}\frac{(G_{\vec{k}+\vec{q}}+G_{-\vec{k}+\vec{q}})G_{\vec{k}'+\vec{q}}}{\bigl[1+U\chi_{GG}^{}(\vec{q})\bigr]\bigl[1-U\chi_{G\bar{G}}^{}(\vec{q})\bigr]}.
\label{Tc-FLEX-S-singlet}
\end{align}
This condition for $T_{\rm c}$, which is also identical with the one that the superconducting correlation function diverges,
is manifestly different from $U\chi_{G\bar{G}}^{}(\vec{0})\!=\!1$.
The same statement holds true in the case of ferromagnetism with $U\!>\!0$, as can be examined easily on the basis of the consideration
of incorporating the spin polarization in Green's function.
Hence, we should allow for the possibility that $U\chi_{G\bar{G}}^{}(\vec{0})\!=\!1$ and $U\chi_{GG}^{}(\vec{0})\!=\!1$ are reached independently
of the phase transitions with broken symmetries.
The fact that $U\chi_{G_0\bar{G}_0}^{}(\vec{0})\!=\!1$ and $U\chi_{G_0G_0}^{}(\vec{0})\!=\!1$ specify
the mean-field $T_{\rm c}$ suggests that
the renormalized ones are realized near the phase transition points with broken symmetries.

Another key observation here is that the values of $U\chi_{G\bar{G}}^{}(\vec{0})$ and $U\chi_{GG}^{}(\vec{0})$ cannot exceed 1 {\it physically}.
This can be seen as follows. The zero Matsubara frequency contribution to the second term of Eq.\ (\ref{Phi_n}) can be written as
\begin{align}
\frac{1}{\beta}\sum_{\bm q} \left\{\ln \left[1-U\chi_{G\bar{G}}(q,0)\right] +U\chi_{G\bar{G}}^{}(q,0)\right\},
\label{Phi-singular}
\end{align}
where $U\chi_{G\bar{G}}(q,0)$ is real, as can be shown from Eq.\ (\ref{chi_AB}).
Since Eq.\ (\ref{Phi-singular}) is part of the grand thermodynamic potential that takes a real value,
the inequality $1-U\chi_{G\bar{G}}^{}(q,0)\!\geq \!1$ should hold.
In other words, the series expansion of Eq.\ (\ref{Phi-singular}) does not converge for $U\chi_{G\bar{G}}^{}(q,0)>1$.
Thus, we are led to the conclusion that $U\chi_{G\bar{G}}^{}(\vec{0})$ and $U\chi_{GG}^{}(\vec{0})$ must have the common bound 1,
at which they should be pinned after reaching it.
The statement holds true not only for $q\!=\!0$ of isotropic systems but also for any value of ${\bm q}$
even in anisotropic systems;
we call  $U\chi_{G\bar{G}}^{}({\bm q},0)\!=\!1$ and $U\chi_{GG}^{}({\bm q},0)\!=\!1$ {\it hidden singularities}. 
It is shown in Appendix\ref{AppA} that
the hidden singularities can be defined generally beyond the FLEX-S approximation.

However, the inequalities $U\chi_{G\bar{G}}^{}(\vec{0})\!\leq \!1$ and $U\chi_{GG}^{}(\vec{0})\!\leq \!1$ cannot be satisfied automatically by the standard Luttinger-Ward formalism based on Eqs.\ (\ref{Omega}) and (\ref{SD}). 
Specifically, our numerical study clarifies that 
we cannot obtain any convergent solution 
below the temperature at which $U\chi_{G\bar{G}}^{}(\vec{0})\!=\!1$ is realized;
there, the self-consistent equations naturally go into the region of 
$U\chi_{G\bar{G}}(0)\!>\!1$, where 
there is a value $q\!=\!q_0^{}\!>\!0$ 
around which $1\!-\!U\chi_{G\bar{G}}^{}(q,0)$ vanishes linearly so that the $q$ integral of Eq.\ (\ref{Sigma_n}) diverges,
thereby making $\Sigma(\vec{k})$ diverge for any value of $\vec{k}$
and $G(\vec{k})$ vanish for any value of $\vec{k}$ accordingly.
The statement holds true even in the presence of lattice anisotropy,
as shown easily by dividing the ${\bm q}$ integrations into those parallel and perpendicular to the plane $U\chi_{G\bar{G}}^{}({\bm q},0)\!=\!{\it const}$.
To the best of our knowledge, neither the hidden singularities nor the corresponding difficulty 
in the Luttinger-Ward formalism have been discussed thus far.

We propose a resolution to this issue
within the variational framework of Eqs.\ (\ref{Omega}) and (\ref{SD})
by using the method of the Lagrange multiplier so as not to affect $\Omega$ explicitly.
Specifically for the negative-$U$ case where $U\chi_{G\bar{G}}^{}(\vec{0})$ is relevant, 
we add to the $\Phi$ functional the following term that vanishes for $U\chi_{G\bar{G}}^{}(\vec{0})=1$:
\begin{align}
\varDelta\Phi_\lambda^{\rm n} \equiv \lambda_{\rm n}^{}N\bigl[1-U\chi_{G\bar{G}}^{}(\vec{0})\bigr],
\label{Phi_lambda^n}
\end{align}
where the number of particles, $N$, is incorporated in the prefactor to make the Lagrange multiplier $\lambda_{\rm n}^{}$
intensive; note that $\lambda_{\rm n}^{}$ has the dimension of energy.
Performing the differentiation of Eq.\ (\ref{SD}) with Eq.\ (\ref{Phi_lambda^n}) added to $\Phi$, 
we obtain the expression of the self-energy for the pinned phase of $U\chi_{G\bar{G}}^{}(\vec{0})=1$ 
in terms of Eq.\ (\ref{Sigma_n}) as
\begin{align}
\Sigma_{\lambda}^{\rm n}(\vec{k}) =\Sigma_{\rm n}(\vec{k})+\bar{n}U\lambda_{\rm n}^{} G(-\vec{k}) 
\label{dSigma_lambda}
\end{align}
with $\bar{n}\!\equiv\! N/V$.
The additional second term stemming from Eq.\ (\ref{Phi_lambda^n})
is characteristic of the pinned phase, which corresponds to the condensate fraction of the ideal Bose-Einstein condensation \cite{Leggett06}.
It has an unusual structure called {\it one-particle reducible} (1PR), i.e., the structure that becomes disconnected upon cutting a line,\cite{CJT74}
which originates from the two-particle reducible structure of Eq.\ (\ref{Phi_lambda^n}).
The parameter $\lambda_{\rm n}^{}$ in Eq.\ (\ref{Phi_lambda^n}) enables us to control the value of $U\chi_{G\bar{G}}^{}(\vec{0})$
and should be chosen in such a way that $U\chi_{G\bar{G}}^{}(\vec{0})\!=\!1$ is met by self-consistently solving the Dyson 
equation:
\begin{align}
G(\vec{k})\!=\!\bigl[G_0^{-1}(\vec{k})\!-\! \Sigma_{\lambda}^{\rm n}(\vec{k})\bigr]^{-1} .
\label{Dyson}
\end{align}
This completes the formulation for the normal state.

The parameter $\lambda_{\rm n}$ is finite only for the pinned phase of $U\chi_{G\bar{G}}^{}(\vec{0})\!=\!1$
and is expected to grow almost linearly as a function of temperature initially.
This linear change in the structure of the self-energy need not be considered in performing the first derivatives of $\Omega$
because of stationarity condition (\ref{SD}).
However, it does bring singularities in the second derivatives of $\Omega$
such as specific heat, similarly to those caused by the emergences of $\Delta(\vec{k})$ for superconductivity and the spin-polarized self-energy for ferromagnetism  near $T_{\rm c}$.
Thus, the point $U\chi_{G\bar{G}}^{}(\vec{0})\!=\!1$ may be regarded as specifying a kind of transition
without apparent broken symmetries. 
In this respect, note that Green's function with the self-energy of Eq.\ (\ref{dSigma_lambda})
cannot be approximated by the quasiparticle form of the Fermi-liquid theory,\cite{AGD63} $G(\vec{k})\approx a/(i\varepsilon_n-\xi_{\bm k})$; 
hence, the singular point may be identified as a non-Fermi-liquid transition.
Detailed studies on observable singularities remain to be performed in the future.

Several comments are in order on the hidden singularities and also on Eq.\ (\ref{Phi_lambda^n})
introduced here to treat them.
There are two possibilities for the hidden singularities.
The first possibility is that the singularities specify the limits of the Luttinger-Ward formalism beyond which it cannot be used, with
no real physical phenomena connected to them; the statement implies that there is a large domain, especially in and near ordered phases with broken symmetries, where the self-consistent perturbative scheme is ineffective.
The second possibility is that the singularities exist physically, as assumed in the bare perturbation scheme with the names of Thouless and Stoner criteria, which should be handled appropriately 
once they have been reached, either by extending the Luttinger-Ward formalism itself or by
relying on some alternative formalism suitable for strongly correlated systems;
since the hidden singularities are generally distinct from the second-order transition points 
with broken symmetries, critical fluctuations should be irrelevant to resolving the issue.
We have taken the second possibility and developed the method given by Eq.\ (\ref{Phi_lambda^n})
to take care of the singularities within the Luttinger-Ward formalism,
through our numerical calculations for the negative-$U$ Hubbard model elaborated below in Sect.\ 3.
This method yields a definite theoretical prediction, which can be tested by experiments, that
the reaching a singularity can be detected as singularities in the second derivatives of the thermodynamic potential
such as specific heat, as mentioned above.
Moreover, the resulting Eq.\ (\ref{dSigma_lambda}) with $G(-\vec{k})$ naturally explains 
the emergence of a pole in the self-energy, which has been discussed 
in terms of the pseudogap behaviors in high-$T_{\rm c}$ cuprates phenomenologically\cite{Norman07}
and also in the negative-$U$ Hubbard model
found numerically by a completely different approach of the dynamical mean-field theory.\cite{PB15,Sakai15}
These are some of the plausible features of our approach with Eq.\ (\ref{Phi_lambda^n}).

\subsection{Superconducting state}

The above consideration for the normal state can be extended directly and continuously to the $s$-wave superconducting state
characterized by the emergence of the anomalous Green's function $F(\vec{k})\!=\!F(-\vec{k})$.
Specifically, we can construct the $\Phi$ functional in the FLEX-S approximation 
so as to incorporate all the anomalous processes derivable from those of Eq.\ (\ref{Phi_n}) 
based on the diagrammatic expression in Fig.\ \ref{fig1} \cite{Kita11},
\begin{align}
\Phi 
 =&\, -\frac{U}{\beta}\sum_{\vec{q}}\chi_{+}^{}(\vec{q}\,)-\frac{1}{2\beta} \sum_{\vec{q}}\bigl[U\chi_+^{}(\vec{q}\,)\bigr]^2
\notag \\
&\, +\frac{1}{2\beta}\sum_{\vec{q}}{\rm Tr}\,\biggl[\ln\bigl(\underline{1}+\underline{x}\bigr)-\underline{x}+\frac{1}{2}\underline{x}^2\biggr]_{\underline{x}=U\underline{\chi}^{({\rm c})}(\vec{q}\,)}
\notag \\
&\, +\frac{3}{2\beta}\sum_{\vec{q}}\biggl[\ln\bigl(1-x\bigr)+x+\frac{1}{2}x^2\biggr]_{x=U\chi_+^{}(\vec{q}\,)}.
\label{Phi}
\end{align}
Here $\underline{1}$ is the $3\times 3$ unit matrix, $\underline{\chi}^{({\rm c})}(\vec{q}\,)$ is
defined in terms of Eq.\ (\ref{chi_AB}) with $A,B\!=\! G,\bar{G},F$ by 
\begin{subequations}
\begin{align}
\underline{\chi}^{({\rm c})}(\vec{q}\,)\equiv 
\begin{bmatrix}
\vspace{1mm}
\chi_{-}^{}(\vec{q}\,)  & \sqrt{2}\chi_{GF}^{}(\vec{q}\,) & -\sqrt{2}\chi_{\bar{G}F}^{}(\vec{q}\,)\\
\vspace{1mm}
\sqrt{2}\chi_{GF}^{}(\vec{q}\,) & -\chi_{G\bar{G}}^{}(\vec{q}\,) & -\chi_{FF}^{}(\vec{q}\,) \\
-\sqrt{2}\chi_{\bar{G}F}^{}(\vec{q}\,) & -\chi_{FF}^{}(\vec{q}\,) & -\chi_{\bar{G}G}^{}(\vec{q}\,) 
\end{bmatrix},
\label{chi^(0c)}
\end{align}
and $\chi_\pm^{}(\vec{q}\,)$ denote
\begin{align}
\chi_\pm^{}(\vec{q}\,)\equiv &\,\chi_{GG}^{}(\vec{q}\,)\pm \chi_{FF}^{}(\vec{q}\,) .
\end{align}
\end{subequations}
The matrix $\underline{\chi}^{({\rm c})}(\vec{q}\,)$ represents the mixing of superconducting and charge fluctuations
due to the emergence of $F$.
Equation (\ref{Phi_lambda^n}) is now replaced by
\begin{align}
\varDelta\Phi_\lambda=\lambda N\det \bigl[\underline{1}+U\underline{\chi}^{({\rm c})}(\vec{q}\,)\bigr] ,
\label{Phi_lambda}
\end{align}
where $\lambda$ is determined by the condition 
that the smallest eigenvalue of the matrix $\underline{1}+U\underline{\chi}^{({\rm c})}(\vec{q}\,)$
stays zero once this has been reached.
Although it should be removed when all the eigenvalues are positive, our numerical study given below in Sect.\ \ref{sec:NR} presents an example 
that incorporating $\varDelta\Phi_\lambda$ is indispensable to obtaining physical (i.e., finite) solutions throughout the superconducting phase.

The contributions of Eq.\ (\ref{Phi}) to the normal and anomalous self-energies can be determined by calculating
$\Sigma(\vec{k})\!=\!\frac{\beta}{2}\frac{\delta\Phi}{\delta G(\vec{k})}$ and $\Delta(\vec{k})\!=\!\frac{\beta}{2}\frac{\delta\Phi}{\delta F(\vec{k})}$,
yielding \cite{Kita11}
\begin{subequations}
\begin{align}
\Sigma(\vec{k})  
 = &\, \frac{UN}{2V}+\frac{1}{\beta V}\sum_{\vec{q}}U_{22}^{\rm eff}(\vec{q}\,)G(-\vec{k}+\vec{q}\,) 
\notag \\
&\, +\frac{1}{2\beta V}\sum_{\vec{q}}\bigl[3U_+^{\rm eff}(\vec{q}\,)-U_{11}^{\rm eff}(\vec{q}\,)\bigr] G(\vec{k}-\vec{q}\,)
\notag \\
&\, -\frac{\sqrt{2}}{\beta V}\sum_{\vec{q}}U_{12}^{\rm eff}(\vec{q}\,)F(\vec{k}-\vec{q}\,),
\label{Sigma}
\end{align}
\begin{align}
\Delta(\vec{k})  
 = &\, \frac{U}{\beta V}\sum_{\vec{k}'}F(\vec{k}')+\frac{1}{\beta V}\sum_{\vec{q}}U_{23}^{\rm eff}(\vec{q}\,)F(\vec{k}-\vec{q}\,) 
\notag \\
&\, +\frac{1}{2\beta V}\sum_{\vec{q}}\bigl[3U_+^{\rm eff}(\vec{q}\,)+U_{11}^{\rm eff}(\vec{q}\,)\bigr] F(\vec{k}-\vec{q}\,)
\notag \\
&\, -\frac{1}{\beta V}\sum_{\vec{q}}U_{12}^{\rm eff}(\vec{q}\,) \frac{G(\vec{k}+\vec{q}\,)+G(-\vec{k}+\vec{q}\,)}{\sqrt{2}} .
\label{Delta}
\end{align}
\end{subequations}
Here, $U_+^{\rm eff}(\vec{q}\,)$ is defined by
\begin{subequations}
\begin{align}
U_+^{\rm eff}(\vec{q}\,)\equiv U\biggl[\frac{U\chi_{+}(\vec{q}\,)}{1-U\chi_{+}(\vec{q}\,)}-\frac{1}{3}U\chi_{+}(\vec{q}\,)\biggr]
\end{align}
and $U_{ij}^{\rm eff}(\vec{q}\,)$ denotes the $ij$ element of the matrix
\begin{align}
\underline{U}^{\rm eff}(\vec{q}\,)\equiv U\bigl[U\underline{\chi}^{({\rm c})}(\vec{q}\,)\bigr]^2\bigl[\underline{1}+U\underline{\chi}^{({\rm c})}(\vec{q}\,)\bigr]^{-1}.
\end{align}
\end{subequations}
After adding Eq.\ (\ref{Phi_lambda}) to the $\Phi$ functional,
the self-energies acquire terms proportional to Green's functions as
\begin{subequations}
\label{SigmaDelta_lambda-s}
\begin{align}
\Sigma_\lambda(\vec{k})=&\, \Sigma(\vec{k})+\bar{n}U\!\left[\lambda_{\bar{G}}^{}G(-\vec{k})+\lambda_{G}^{}G(\vec{k}) + \lambda_{F}^{}F(\vec{k})\right],
\label{Sigma_lambda-s}
\\
\Delta_\lambda(\vec{k})=&\, \Delta(\vec{k})+ \bar{n}U\!\left\{\tilde\lambda_{F}^{}F(\vec{k})+\tilde\lambda_{G}^{}[G(\vec{k})+ G(-\vec{k})]\right\},
\label{Delta_lambda-s}
\end{align}
\end{subequations}
with 
\begin{align*}
\lambda_{\bar{G}}^{}\!\equiv &\,2\lambda\bigl\{\bigl[1\!+\!U\chi_{-}^{}(\vec{0})\bigr]\bigl[1\!-\!U\chi_{G\bar{G}}^{}(\vec{0})\bigr]
\!-\!2\bigl[U\chi_{GF}^{}(\vec{0})\bigr]^2\bigr\},
\\
\lambda_{G}^{}\!\equiv &\, -\lambda\bigl\{\bigl[1\!-\!U\chi_{G\bar{G}}^{}(\vec{0})\bigr]^2
\!-\!\bigl[U\chi_{FF}^{}(\vec{0})\bigr]^2\bigr\},
\\
\lambda_{F}^{}\!\equiv &\, 4\lambda\, U\chi_{GF}^{}(\vec{0})\bigl[1\!-\!U\chi_{G\bar{G}}^{}(\vec{0})\!-\!U\chi_{FF}^{}(\vec{0})\bigr],
\\
\tilde\lambda_{F}^{}\!\equiv &\,2\lambda\bigl\{\bigl[1\!+\!U\chi_{-}^{}(\vec{0})\bigr]U\chi_{FF}^{}(\vec{0})
\!-\!2\bigl[U\chi_{GF}^{}(\vec{0})\bigr]^2\bigr\}
\\
&\, +\lambda\bigl\{\bigl[1\!-\!U\chi_{G\bar{G}}^{}(\vec{0})\bigr]^2
\!-\!\bigl[U\chi_{FF}^{}(\vec{0})\bigr]^2\bigr\},
\\
\tilde\lambda_{G}^{}\!\equiv &\, 2\lambda\, U\chi_{GF}^{}(\vec{0})\bigl[1\!-\!U\chi_{G\bar{G}}^{}(\vec{0})
\!-\!U\chi_{FF}^{}(\vec{0})\bigr].
\end{align*} 
The Dyson-Gor'kov equation can be written in terms of these self-energies as
\begin{align}
\begin{bmatrix}
G(\vec{k}) \\ F(\vec{k})
\end{bmatrix}=\frac{1}{D(\vec{k})}\begin{bmatrix}
-G_0^{-1}(-\vec{k})+\Sigma_\lambda(-\vec{k}) \\ \Delta_\lambda(\vec{k})
\end{bmatrix},
\label{DG}
\end{align}
with $D(\vec{k})\!\equiv\! [G_0^{-1}(\vec{k})\!-\!\Sigma_\lambda(\vec{k})][-G_0^{-1}(-\vec{k})\!+\!\Sigma_\lambda(-\vec{k})]\!-\![\Delta_\lambda(\vec{k})]^2$.
It has turned out numerically that $\lambda_{\bar{G}}\!=\!\tilde\lambda_{F}^{}\!>\!0$ and $\lambda_{G}^{}\!=\!\lambda_{F}^{}\!=\!\tilde\lambda_{G}^{}\!=\!0$ hold
for $\lambda$
that meets $\det[\underline{1}+U\underline{\chi}^{({\rm c})}(\vec{q}\,)]=0$.
Note that $\lambda_{\bar{G}}^{}$ should reach $\lambda_{\rm n}^{}$ of Eq.\ (\ref{dSigma_lambda}) 
at the superconducting transition temperature.
This completes our formulation to study the superconducting phase continuously from the normal state
on the basis of the FLEX-S approximation.

\section{Numerical Results\label{sec:NR}}

\begin{figure}[b]
\includegraphics[width=0.8\linewidth]{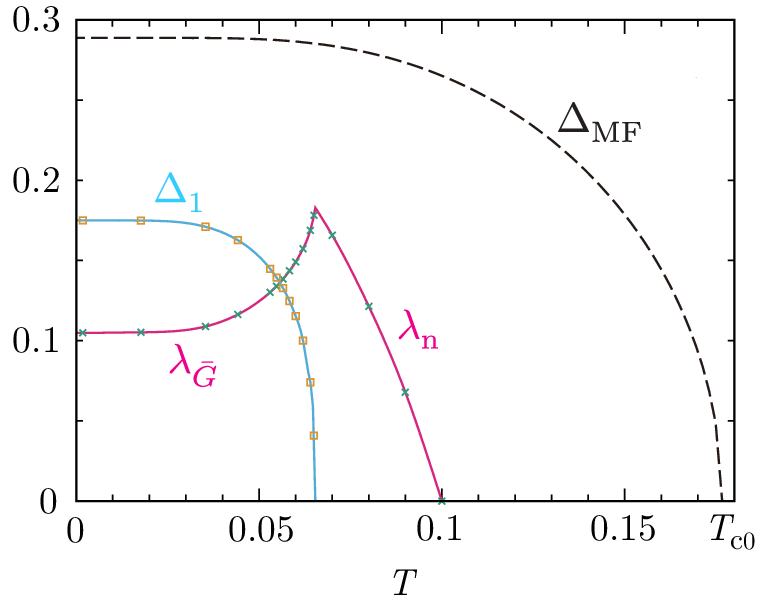}
\caption{\label{fig2}Temperature dependences of the Lagrange multipliers $(\lambda_{\rm n}^{},\lambda_{\bar{G}}^{})$
and the first-order anomalous self-energy $\Delta_1^{}$ in comparison with that of the mean-field energy gap $\Delta_{\rm MF}^{}$.
}
\end{figure}

Self-consistent calculations of the Dyson and Dyson-Gor'kov equations have been performed numerically on the negative-$U$ Hubbard model at a low electron density under the constant density
\begin{align}
\frac{N}{V}=\frac{2}{\beta V}\sum_{\vec{k}} G(\vec{k})\,e^{i\varepsilon_n0_+} ,
\label{mu-eq}
\end{align}
where $0_+$ denotes an infinitesimal positive constant; the technical details are given in Appendix\ref{AppB}.
We have adopted the units of $2m\!=\!\hbar\!=\!k_{\rm F}\!=\!1$, where $m$ is the electron mass and $k_{\rm F}$ is the Fermi momentum.
The momentum cutoff $k_{\rm c}\!=\!10$ has been introduced to describe a finite band width.
Specifically, our $G_0^{-1}(\vec{k})$ can be written as $G_0^{-1}(\vec{k})\!=\!i\varepsilon_n\!-\!k^2\!+\!\mu$ with  $0\!\leq \!k\!\leq\! 10$, 
where $\mu$ is the chemical potential determined by Eq.\ (\ref{mu-eq}).
We have chosen the parameter $U$ as $UN(0)=-0.09$ in terms of the density of states per spin and the volume at the Fermi energy
$N(0)\!=\! \frac{1}{4\pi^2}$; note also that $\frac{N}{V}\!=\!\frac{1}{3\pi^2}$ in the present units.
The corresponding mean-field superconducting transition temperature $T_{{\rm c}0}$,
which is evaluated by using only the first terms on the right-hand sides
of Eqs.\ (\ref{Sigma}) and (\ref{Delta}),
is $0.1767$ that is one order of magnitude smaller than the noninteracting Fermi energy $\varepsilon_{\rm F}^0\!=\! 1$.
Thus, our system lies in the intermediate coupling regime.

Figure \ref{fig2} shows the temperature dependences of the Lagrange multipliers $\lambda_{\rm n}^{}$ and $\lambda_{\bar{G}}$ of Eqs.\ (\ref{dSigma_lambda}) and (\ref{Sigma_lambda-s}), respectively, together with
 the first-order anomalous self-energy $\Delta_1^{}$ defined by the first term 
on the right-hand side of Eq.\ (\ref{Delta}) in terms of the fully self-consistent $F$ with correlations,
which is to be compared with the mean-field energy gap $\Delta_{\rm MF}^{}$.
The crosses and squares denote numerical results, and smooth curves are obtained by interpolating them.
Matching the curves of $\lambda_{\rm n}^{}$ and $\lambda_{\bar{G}}$ yields an estimate of the superconducting transition temperature
$T_{{\rm c}}^{}\!=\! 0.0654$ with correlations, 
which is far below the mean-field value $T_{{\rm c}0}\!=\!0.1767$.
This mean-field calculation yields the ratio $\Delta_{\rm MF}^{}(0)/T_{{\rm c}0}^{}\!=\! 1.635$ that is still close to the weak-coupling value of
$1.76$ \cite{Leggett06}, whereas that with correlations evaluated by $\Delta_1^{}(0)/T_{{\rm c}}^{}$ is increased up to $2.68$.
Note that all the quantities in Fig.\ \ref{fig2}, which originally have the dimension of energy,
are of the same order as $|UN(0)|=0.09$, as may be expected naturally.

\begin{figure}[b]
\includegraphics[width=0.8\linewidth]{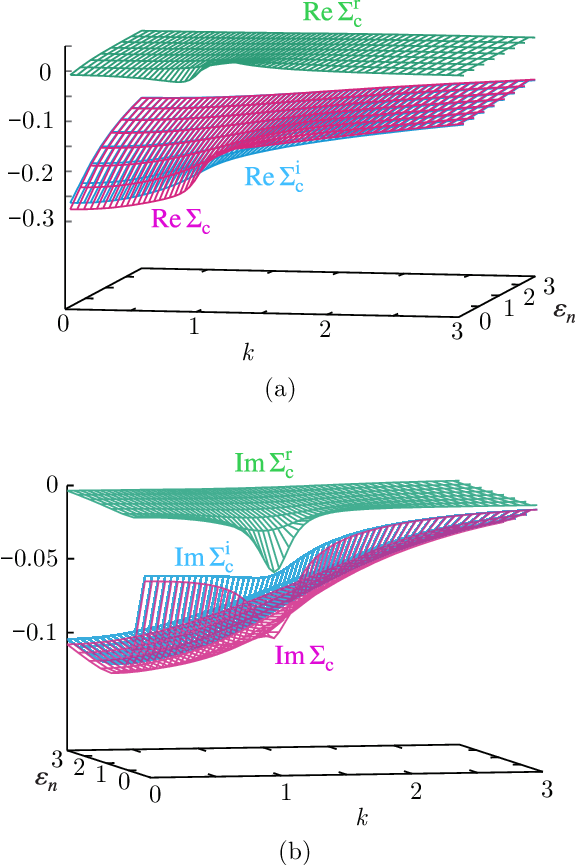}
\caption{\label{fig3}(a) Real and (b) imaginary parts of the correlation self-energy $\Sigma_{\rm c}(k,i\varepsilon_n^{})$ 
calculated at $T\!=\!0.08$, which is expressed as the sum $\Sigma_{\rm c}\!=\! \Sigma_{\rm c}^{\rm i}\!+\!\Sigma_{\rm c}^{\rm r}$
of the 1PI and 1PR contributions $\Sigma_{\rm c}^{\rm i}$  and $\Sigma_{\rm c}^{\rm r}$, respectively.
}
\end{figure}

One of the key features of Fig.\ \ref{fig2} is that $\lambda_{\rm n}^{}$ starts to develop well above $T_{\rm c}^{}\!=\! 0.0654$
at $T_{1{\rm PR}}^{}\!=\! 0.100$,
which means that the singularity of $U\chi_{G\bar{G}}^{}(0)\!=\! 1$ is reached already in the normal state.
This is a continuous transition with no apparent broken symmetries, which is characterized by the emergence of 
the 1PR structure in the self-energy. 
Figure \ref{fig3} shows the (a) real and (b) imaginary parts of the correlation self-energy $\Sigma_{{\rm c}}$,
i.e., Eq.\ (\ref{dSigma_lambda}) without the $\frac{UN}{2V}$ term in Eq.\ (\ref{Sigma_n}),  calculated at $T\!=\!0.08$,
where $\Sigma_{{\rm c}}^{\rm i}$ and $\Sigma_{{\rm c}}^{\rm r}$ denote the one-particle-irreducible (1PI) and 1PR 
contributions, respectively.
Thus, the emergence of $\Sigma_{{\rm c}}^{\rm r}$ brings a marked change in the structure of the self-energy,
especially around the Fermi momentum at low Matsubara frequencies, as expected naturally from
$\Sigma_{{\rm c}}^{\rm r}(\vec{k})\!\propto\! G(-\vec{k})$.

\begin{figure}[t]
\includegraphics[width=0.9\linewidth]{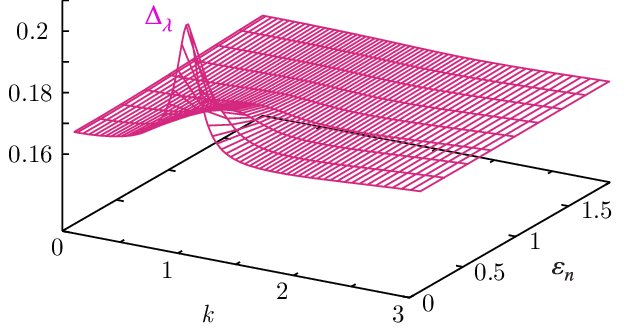}
\caption{\label{fig4}Anomalous self-energy $\Delta_\lambda(k,i\varepsilon_n)$ calculated at $T\!=\! 0.0353$.
}
\end{figure}
The 1PR structure is found to persist below $T_{1{\rm PR}}^{}$ into the superconducting phase
and emerges also in the anomalous self-energy $\Delta_\lambda^{}$.
As seen in Fig.\ \ref{fig4} calculated at $T\!=\! 0.2T_{{\rm c}0}\!=\!0.0353$, 
there is a conspicuous peak in the background of $\Delta_1^{}\!\sim\! 0.17$ 
around the Fermi momentum $k\!=\! 1$ at low Matsubara frequencies,
which originates from the 1PR contribution $\Delta_{\rm c}^{\rm r}(\vec{k})\!\propto\! F(\vec{k})$.

\begin{figure}[t]
\includegraphics[width=0.8\linewidth]{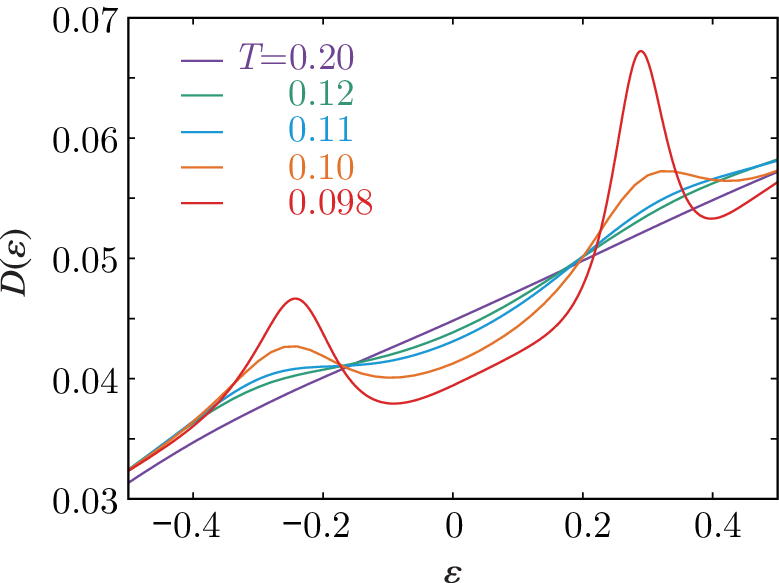}
\caption{\label{fig5}The density of states around the excitation threshold calculated at five different temperatures.}
\end{figure}

The emergence of the 1PR structure in the self-energy is necessarily accompanied by a change in the single-particle density of states
per volume,
\begin{align}
D(\varepsilon)\!\equiv \! -\frac{2}{\pi V}\sum_{{\bm k}}{\rm Im}\, G(k,i\varepsilon_n^{}\rightarrow \varepsilon\!+\! i 0_+^{}).
\end{align}
Figure \ref{fig5} shows the density of states over $-0.5\!\leq\!\varepsilon\!\leq 0.5$ at
five different temperatures obtained by a numerical analytic continuation 
of the self-energy based on the six-point Pad\'e approximant \cite{VS77}.
As seen in this figure, the reduction in the density of states around the excitation threshold, called the {\it pseudogap behavior}, starts 
even from above $T_{1{\rm PR}}\!=\!0.100$ and becomes increasingly conspicuous down through $T_{1{\rm PR}}$.
The singular 1PR structure makes a reliable analytic continuation based on the six-point Pad\'e approximant
impossible below $T\!=\! 0.098$, which is an issue to be clarified in the future.
The present mechanism for the pseudogap behavior may explain those observed  in 
ultracold atomic gases.\cite{SGJ08,GSDJPPS10}
In this respect, it is worth noting that the pseudogap behavior has been reproduced 
qualitatively by numerical calculations
on the negative-$U$ Hubbard model based on the dynamical mean-field theory,\cite{PB15,Sakai15} and attributed 
to the emergence of a pole in the self-energy in the strong-coupling low-temperature region;\cite{Sakai15}
the present study can trace the microscopic origin of this pole to the second term of Eq.\ (\ref{dSigma_lambda}).
Since the hidden singularity that we have found is general and may not depend on microscopic details of the model, it is interesting to explore in the future whether the present mechanism can explain the pseudogap observed in cuprate superconductors\cite{TS99,NPK05,Matsuda17} by taking an appropriate ${\bm q}$ for 
either $U\chi_{G\bar{G}}^{}({\bm q},0)\!=\! 1$ or $U\chi_{GG}^{}({\bm q},0)\!=\! 1$.

Note finally that the concept of the 1PR structure in the self-energy has emerged naturally 
in our effort to describe strongly correlated systems
microscopically and numerically; without introducing Eqs.\ (\ref{Phi_lambda^n}) and (\ref{Phi_lambda}), we could not have obtained any finite solution of the Dyson and Dyson-Gor'kov equations 
below $T_{1{\rm PR}}^{}$.
In this respect, our model with a definite cutoff $k_{\rm c}\!=\!10$ may have been advantageous
over the continuous model with ultraviolet divergences \cite{HRCZ07}
in that no further renormalization of $U$, which may mask the singularities, is required.

\section{Conclusions}

We have clarified existence of hidden singularities in correlated electron systems
and developed a definite procedure to handle them.
It is thereby shown that
reaching the singularities is characterized by the emergence of a novel structure in the self-energy 
called 1PR, which is necessarily accompanied by a change in the single-particle density of states 
around the excitation threshold. The temperature at which it starts to develop can be regarded as a transition
with no apparent broken symmetries. 

They singularities are inherent in the interaction channel and 
should also present in Bose systems. They
may have stood as a key untold difficulty in performing theoretical studies on correlated normal and ordered phases
based on the quantum field theory, as may be seen in the negligible number of theoretical studies on them
compared with those on the normal state \cite{Yanase03,Kontani13,Scalapino12}, a resolution for which has been presented here.
The 1PR structure has also been predicted to be present in 
Bose-Einstein condensates on the basis of a self-consistent $\Phi$ derivable scheme of satisfying Goldstone's theorem and conservation laws \cite{Kita09,Kita14}. 
Thus, it may be a common feature over a wide range of correlated systems.
Further detailed studies will be required to fully elucidate the novel phase with the 1PR structure in the self-energy.

\begin{acknowledgments}
The author is grateful to Masatoshi Imada for his valuable comments on the present results.
This work was supported by JSPS KAKENHI Grant Number JP20K03848.
\end{acknowledgments}

\appendix

\section{General Definition of Hidden Singularities\label{AppA}}

\begin{figure}[b]
\includegraphics[width=0.9\linewidth]{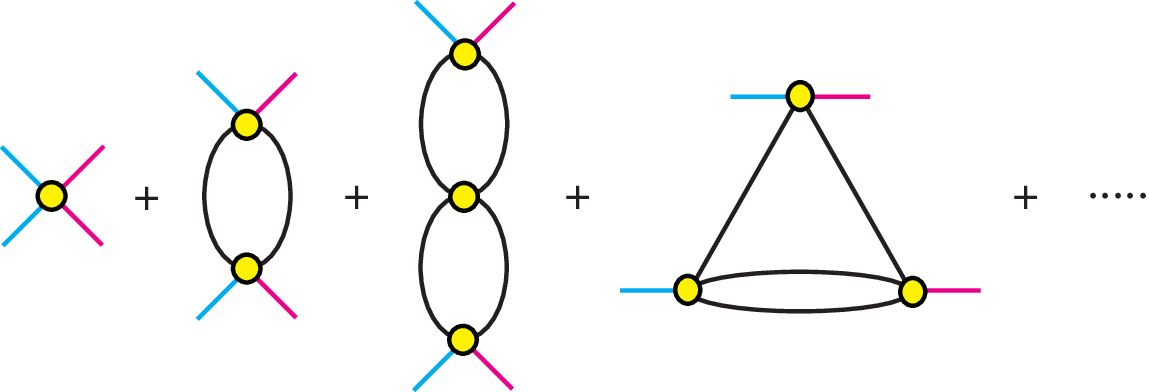}
\caption{\label{figA1}Diagrammatic expression of $\Gamma^{(\rm ir)}$ defined by Eq.\ (\ref{Gamma^irr}).
The left (right) pair of external lines represents $11'_{ii'}$ ($22'_{jj'}$).
}
\end{figure}

Let us define the irreducible vertex by Eq.\ (39) of Ref.\ \onlinecite{Kita11} as
\begin{align}
\Gamma_{ii',jj'}^{(\rm ir)}(11',22')\equiv 2\beta\frac{\delta^2\Phi}{\delta G_{i'i}(1',1)\delta G_{j'j}(2',2)},
\label{Gamma^irr}
\end{align}
where $1$ is defined by $1\!\equiv\! (\xi_1,\tau_1)$ with $\xi_1$ denoting the space-spin coordinates and $\tau_1\!\in\![0,\beta]$,
and the subscript $i\!=\!1,2$ distinguishes the creation and annihilation operators.
The differentiation of Eq.\ (\ref{Gamma^irr}) amounts to cutting a pair of lines in Fig.\ \ref{fig1} in all possible ways. 
Hence, the irreducible vertex can be expressed diagrammatically as shown in Fig.\ \ref{figA1}, which has the structure called {\it simple}\cite{DDM64-2}
or {\it two-particle irreducible}
with respect to  $11'_{ii'}$ and $22'_{jj'}$; see the paragraph below Eq.\ (43) in Ref.\ \onlinecite{DDM64-2}.
We also introduce the matrices $\underline{\Gamma}^{(\rm ir)}$, $\underline{\chi}$, and  $\underline{1}$ as
\begin{subequations}
\label{uGamma-uchi}
\begin{align}
\langle 11'_{ii'}|\underline{\Gamma}^{(\rm ir)}|22'_{jj'}\rangle \equiv &\,
{\Gamma}^{(\rm ir)}_{ii',j'j}(11',2'2),
\label{uGamma^irr}
\\
\langle 11'_{ii'}|\underline{\chi}|22'_{jj'}\rangle \equiv &\,
-G_{ij}(1,2)G_{j'i'}(2',1'),
\label{uGG}
\\
\langle 11'_{ii'}|\underline{1}|22'_{jj'}\rangle \equiv &\,
\delta_{ij}\delta_{i'j'}\delta(1,2)\delta(1',2'),
\label{u1}
\end{align}
\end{subequations}
in the same way as Eqs.\ (25) and (42d) in Ref.\ \onlinecite{Kita11}.
Using them, we can generalize Eq.\ (28) in Ref.\ \onlinecite{Kita11} 
in the FLEX-S approximation to obtain the whole logarithmic contribution to $\Phi$:
\begin{align}
\Phi_{\rm log}=\frac{1}{2\beta}{\rm Tr}\ln \bigl(\underline{1}+\underline{\Gamma}^{(\rm ir)}\underline{\chi}\bigr).
\label{Phi_log}
\end{align}
We then transform the $\tau$ variables in the brackets of Eq.\ (\ref{uGamma^irr}) into Matsubara frequencies as
\begin{align}
\langle 11'|\underline{\Gamma}^{(\rm ir)}|22'\rangle=&\,\frac{1}{\beta^3}\sum_{\ell}\sum_{n_1n_2}
\langle\tilde{1}\tilde{1}'_{ii'}|\underline{\tilde\Gamma}^{(\rm ir)}(\omega_\ell)|\tilde{2}\tilde{2}'_{jj'}\rangle
\notag \\
&\,\times e^{-i\varepsilon_1\tau_1 
+i\varepsilon_{1-}\tau_{1'} +i\varepsilon_2\tau_2-i\varepsilon_{2-}\tau_{2'}},
\end{align}
where $\underline{\tilde\Gamma}^{(\rm ir)}$ denotes a matrix with the index
$|\tilde{1}\tilde{1}'_{ii'}\rangle\equiv|\xi_1\xi_{1'},\varepsilon_{n_1},ii'\rangle$, the symbol $\varepsilon_1$ is the abbreviation for
$\varepsilon_{n_1}$,
and $\varepsilon_{1-}$ is defined by $\varepsilon_{1-}\equiv\varepsilon_{n_1}-\omega_\ell$.
The matrices of Eqs.\ (\ref{uGG}) and (\ref{u1}) can be expanded similarly
in terms of the coefficients
\begin{subequations}
\begin{align}
\langle\tilde{1}\tilde{1}'_{ii'}|\underline{\tilde\chi}(\omega_\ell)|\tilde{2}\tilde{2}'_{jj'}\rangle
\equiv &\,-\beta\delta_{n_1n_2}G_{ij}(\xi_1,\xi_2;\varepsilon_1)
\notag \\ &\,\times G_{j'i'}(\xi_{2'},\xi_{1'};\varepsilon_{n_1}\!-\!\omega_\ell),
\end{align}
\begin{align}
\langle\tilde{1}\tilde{1}'_{ii'}|\underline{\tilde{1}}|\tilde{2}\tilde{2}'_{jj'}\rangle
\equiv &\, \beta\delta_{n_1n_2}\delta_{ij}\delta_{i'j'}\delta(\xi_1,\xi_2)\delta(\xi_{1'},\xi_{2'}).
\end{align}
\end{subequations}
Using the tilde matrices, we can express Eq.\ (\ref{Phi_log}) as
\begin{align}
\Phi_{\rm log}=&\,\frac{1}{2\beta}\sum_{\ell}{\rm Tr} \,\ln \Bigl[\underline{\tilde{1}}+\underline{\tilde\Gamma}^{(\rm ir)}(\omega_\ell)\,\underline{\tilde\chi}(\omega_\ell)\Bigr]
\notag \\
=&\, \frac{-1}{2\beta}\sum_{\ell}{\rm Tr} \,\ln \Bigl[\underline{\tilde{1}}-\underline{\tilde\Gamma}(\omega_\ell)\,\underline{\tilde\chi}(\omega_\ell)\Bigr],
\label{Phi_log2}
\end{align}
where $\underline{\tilde\Gamma}$ is the full vertex that obeys $\underline{\tilde\Gamma}\!=\!\underline{\tilde\Gamma}^{(\rm ir)}
-\underline{\tilde\Gamma}^{(\rm ir)}\underline{\tilde\chi}\,\underline{\tilde\Gamma}$.
The hidden singularity is defined generally as the point where the determinant of
$\underline{\tilde{1}}+\underline{\tilde\Gamma}^{(\rm ir)}(0)\,\underline{\tilde\chi}(0)$
reaches zero, or equivalently, the determinant of
$\underline{\tilde{1}}-\underline{\tilde\Gamma}(0)\,\underline{\tilde\chi}(0)$ diverges. 
Reducing the matrix by removing the singularity, one may find multiple singularities.

By either cutting the series in Fig.\ \ref{figA1} at some finite order or taking its partial sum to infinity, we obtain an approximate treatment of the logarithmic singularity
of Eq.\ (\ref{Phi_log2}), such as the FLEX-S approximation of Eqs.\ (\ref{Phi_n}) and (\ref{Phi})
that cuts the series in Fig.\ \ref{figA1} at the lowest order.
Alternatively, the parameter $U$ in Eqs.\ (\ref{Phi_n}) and (\ref{Phi}) may be regarded as an effective potential
for the full irreducible vertex of Eq.\ (\ref{Phi_log2}).

Note finally that the hidden singularities do not specify second-order transition points
with broken symmetries. 
For example, the superconducting transition temperature of the isotropic $s$-wave pairing 
is determined by Eq.\ (\ref{Tc-eq}) in terms of $U^{\rm s}_{\vec{k}\vec{k}'}$ defined generally 
by $U^{\rm s}_{\vec{k}\vec{k}'}\!\equiv\! \frac{\beta}{2}\frac{\delta^2\Phi}{\delta F(\vec{k})\delta F(\vec{k}')}\Bigr|_{F=0}$.

\section{Numerical Procedures\label{AppB}}

Numerical calculations of Eqs.\ (\ref{Dyson}) and (\ref{DG}) have been performed iteratively by choosing the noninteracting and mean-field
Green's functions as the starting points, respectively.
Specifically for the normal state, we divide Green's function 
in the $\tau$ space with $\tau\in[0,\beta]$ as
\begin{align}
G(k,\tau)=G_0(k,\tau)+\delta G(k,\tau), 
\end{align}
where $G_0(k,\tau)$ is the noninteracting Green's function
\begin{align}
G_0(k,\tau)=-\frac{e^{-\tau(k^2-\mu_0)}}{e^{-\beta(k^2-\mu_0)}+1} 
\end{align}
and $\delta G(k,\tau)$ denotes the correlation part updated iteratively starting from $\delta G(k,\tau)\!=\!0$.
The variable $k$ is expressed as
\begin{subequations}
\begin{align}
k=1+\sinh x
\end{align} 
and discretized 
with an equal interval in $x$ over $0\!\leq\! k\!\leq\! k_{\rm c}\!=\!10$; the number of points is $N_k\!=\!50$-$ 200$. 
Similarly, $\tau$ is expressed as 
\begin{align}
\tau=\frac{\beta}{2}(1+\tanh x)
\end{align}
\end{subequations}
and discretized 
with an equal interval in $x$ over $\tau_{\rm c}\!\leq\!\tau\!\leq\! \beta\!-\!\tau_{\rm c}$ with $\tau_{\rm c}\!=\!10^{-8}$;
the number of points is $N_\tau\!=\!50$-$400$.

The correlation functions of Eq.\ (\ref{chi_AB}) are calculated in the $\tau$ space
on the basis of the equations
\begin{subequations}
\label{chi_GG}
\begin{align}
\chi_{GG}^{}(q,\tau)=&\,\frac{1}{V}\sum_{\bm k}G(k,\tau)G(|{\bm k}-{\bm q}|,\beta-\tau),
\\
\chi_{G\bar{G}}^{}(q,\tau)=&\, -\frac{1}{V}\sum_{\bm k}G(k,\tau)G(|{\bm k}-{\bm q}|,\tau).
\end{align}
\end{subequations}
The angular integrals of $s\equiv{\bm k}\!\cdot\!{\bm q}/kq$ in Eq.\ (\ref{chi_GG}) are performed 
over 
\begin{align}
{\max}\!\left(-1,\frac{k^2\!+\!q^2\!-\!k_{\rm c}^2}{2kq}\right)\!\leq s \leq 1
\end{align}
by dividing the interval at $(k^2\!+\!q^2\!-\!1)/2kq$ when the value lies in the interval.
The integration over each interval $s_0\!\leq\! s\!\leq s_1$ is performed by
expressing 
\begin{align}
s=\frac{s_0\!+\!s_1}{2}+\frac{s_0\!-\!s_1}{2}\tanh(\sinh x)
\label{Variable-s}
\end{align}
and discretizing $x$ at an equal interval
over $-2.5\!\leq \! x\!\leq \! 2.5$; the number of integration points is $N_s\!=\! 20$-$30$.
The radial integrals of ${\bm k}$ in Eq.\ (\ref{chi_GG}) are performed similarly 
over
\begin{align}
{\rm max}(0,q-k_{\rm c})\leq k\leq k_{\rm c}
\end{align}
by dividing the interval at the singular points
$|1\!-\!q|$, $1$, $1\!+q$ when they lie in the interval, and changing variables similarly to Eq.\ (\ref{Variable-s}) for each interval.

The variable $q$ of Eq.\ (\ref{chi_GG}) runs over $[0,2k_{\rm c}]$, which is divided at $q\!=\!2,k_{\rm c}$ into three regions.
Each region is then discretized by Eq.\ (\ref{Variable-s}) with $s\rightarrow q$ with the relative number of points 
$4:3:3$; the total number of points is $N_q\!=\!50$-$200$.

The functions of Eq.\ (\ref{chi_GG}) are then transformed into the $\omega_\ell$ space on the basis of the method
developed by Haussmann {\it et al}.\cite{HRCZ07} 
To be specific, we express any function $\chi(\tau)$ with $\tau\!\in\! [\tau_j,\tau_{j+1})$ through the cubic spline,\cite{NRC}
\begin{align}
\chi(\tau) \approx \chi_j(\tau)\equiv\sum_{n=0}^3 a_{j,n}(\tau-\tau_j)^n ,
\label{chi(tau)}
\end{align}
and carry out the Fourier transform as
\begin{align}
\bar\chi(\omega_\ell)\equiv&\, \int_{0}^{\beta}\chi(\tau)\,e^{i\omega_\ell\tau}\, d\tau\approx \sum_{j=1}^{N_{\tau}-1} \int_{\tau_j}^{\tau_{j+1}}\chi_j(\tau)\,e^{i\omega_\ell\tau}\, d\tau
\notag \\
=&\,\sum_{n=0}^3 a_{j,n} I_{j,n} (\omega_\ell),
\label{chi(tau->omega)}
\end{align}
where $I_{j,n}(\omega_\ell)$ is defined by
\begin{align}
I_{j,n}(\omega_\ell)=&\,e^{i\omega_\ell\tau_j}\frac{d^n}{d(i\omega_\ell)^n}\int_{\tau_j}^{\tau_{j+1}}e^{i\omega_\ell\tau}\, d\tau
\notag\\
=&\,e^{i\omega_\ell\tau_j}\sum_{\nu=0}^n \frac{n!}{\nu!}(-1)^{n-\nu} \frac{(\varDelta \tau_j)^\nu\,e^{i\omega_\ell\varDelta \tau_j}-\delta_{\nu 0}}{(i\omega_\ell)^{n-\nu+1}},
\label{I_j,n}
\end{align}
with $\varDelta \tau_j\!\equiv\! \tau_{j+1}\!-\!\tau_j$.
For $|\omega_\ell\varDelta \tau_j|\!<\!0.1$, this function is evaluated by expanding the formula in the Taylor series of $\omega_\ell\varDelta \tau_j$.

The boson Matsubara frequencies $\omega_\ell\!=\!2\pi \ell T$ ($\ell\!=\!0,1,2,\cdots$) for $\ell\!>\! \ell_{\rm c}\!\sim \! 10$ are chosen so that they are
described approximately by $\omega_\ell\!\approx\!\sinh x_j$ in terms of equally spaced points $x_j$ ($j\!=\!\ell_{\rm c}\!+\!2,\cdots,N_\tau$) up to $\omega_\ell \!\approx\!10^7$; the number of points is $N_\omega\!=\! 50$-$200$.
The function $\bar\chi(\omega)$ with $\omega\!\in\! [\omega_{\ell_j}^{},\omega_{\ell_{j+1}}^{})$ is then approximated by the cubic spline as
\begin{align}
\bar\chi(\omega)\approx &\,\bar\chi_j(\omega)\equiv \sum_{n=0}^3 \bar{a}_{j,n}(\omega-\omega_{\ell_j}^{}).
\end{align}

The accuracy of Eq.\ (\ref{chi(tau->omega)}) can be examined by 
the inverse transform 
\begin{align}
\chi(\tau)=&\,\frac{1}{\beta}\sum_\ell \bar{\chi}(\omega_\ell)\,e^{-i\omega_\ell\tau}
\notag \\
\approx &\,\frac{1}{\beta}\bar{\chi}(0) +2{\rm Re} \sum_{j=2}^{N_\omega-1} \bar{a}_{j,n}\bar{I}_{j,n}(\tau) ,
\label{chi(tau)-2}
\end{align}
where $\bar{I}_{j,n}(\tau)$ is defined by
\begin{subequations}
\begin{align}
\bar{I}_{j,n}(\tau)\equiv &\,\frac{1}{\beta}\sum_{\ell=\ell_j}^{\ell_{j+1}-1} (\omega_\ell-\omega_{\ell_j}^{})^n \,e^{-i\omega_\ell\tau}
\notag \\
=&\, \frac{e^{-i\omega_{\ell_j}^{}\tau}}{\beta}\frac{d^n}{d(-i\tau)^n}\frac{e^{-i\varDelta\omega_j\tau}-1}{e^{-i\omega_1\tau}-1},
\label{barI}
\end{align}
with $\varDelta\omega_j\!\equiv \!2\pi (\ell_{j+1}-\ell_j)T$; the upper limit $\ell_{j+1}\!-\!1$ of Eq.\ (\ref{barI}) for $j\!=\!N_\omega\!-\! 1$ should be replaced 
by $\ell_{j+1}$.
When $\varDelta\omega_j\tau<0.1$, we expand the last formula in $\varDelta\omega_j\tau$ as
\begin{align}
\bar{I}_{j,n}(\tau)
=&\, \frac{e^{-i\omega_{\ell_j}^{}\tau}}{\beta}\omega_1^n 
\sum_{\nu=0}^\infty (-1)^\nu \frac{(\tau\omega_1)^{2\nu}}{(2\nu)!}
\biggl[S_{n+2\nu}(L)
\notag \\
&\,-\frac{i\tau\omega_1}{2\nu+1}S_{n+2\nu+1}(L)\biggr],
\label{barI2}
\end{align}
\end{subequations}
with $L\!\equiv\!\ell_{j+1}\!-\!\ell_j\!-\! 1$ and 
\begin{align}
S_\nu(L)\equiv \sum_{\ell=0}^L \ell^\nu ,
\label{S_nu(L)}
\end{align}
and replace the upper limit $\infty$ in Eq.\ (\ref{barI2}) by $k_{\rm c}\!\sim\! 10$ to obtain sufficient accuracy.
The sum of Eq.\ (\ref{S_nu(L)}) for $\nu\!\lesssim \! 10$ can be evaluated using Faulhaber's formula.
Equation (\ref{chi(tau)-2}) after the double Fourier transforms thereby obtained has been confirmed to reproduce Eq.\ (\ref{chi(tau)}) excellently.

The functions $\chi_{GG}^{}(\vec{q}\,)$ and $\chi_{G\bar{G}}^{}(\vec{q}\,)$ calculated from Eq.\ (\ref{chi_GG}) using
Eq.\ (\ref{chi(tau->omega)}) are then substituted in Eq.\ (\ref{Sigma_n}).
The sum over $\ell$ is calculated in the $\tau$ space by the Fourier transforms of
\begin{subequations}
\label{U_phpp}
\begin{align}
U_{\rm ph}(\vec{q}\,)\equiv &\, U\Biggl\{U\chi_{GG}^{}(\vec{q}\,)+\frac{3}{2}\,\frac{[U\chi_{GG}^{}(\vec{q}\,)]^2}{1-U\chi_{GG}^{}(\vec{q}\,)}
\notag\\
&\,-\frac{1}{2}\,\frac{\bigl[U\chi_{GG}^{}(\vec{q}\,)\bigr]^2}{1+U\chi_{GG}^{}(\vec{q}\,)}
\Biggr\},
\label{U_ph}
\\
U_{\rm pp}(\vec{q}\,)\equiv &\,U\frac{[U\chi_{G\bar{G}}^{}(\vec{q}\,)]^2}{1-U\chi_{G\bar{G}}^{}(\vec{q}\,)},
\label{U_pp}
\end{align}
\end{subequations}
as
\begin{align}
\Sigma_{\rm c}(k,\tau)
=&\,\frac{1}{V} \sum_{{\bm q}} \bigl[U_{\rm ph}(q,\tau)G(|{\bm k}-{\bm q}|,\tau)
\notag \\
&\,-U_{\rm pp}(q,\tau)G(|{\bm k}-{\bm q}|,\beta-\tau)\bigr],
\label{Sigma(k,tau)}
\end{align}
where the subscript of $\Sigma_{\rm c}$ denotes the correlation part.
On the other hand, the angular and radial integrals of ${\bm q}$ in Eq.\ (\ref{Sigma(k,tau)})
are carried out in the same way as those of ${\bm k}$ in Eq.\ (\ref{chi_GG}). 
Possible singular points 
in the radial integral of $q\!\in\![0,k\!+\!k_{\rm c}]$ are $|1\!-\! k|$, $1\!+\! k$, and $k_{\rm c}\!-\!k$.

Equation (\ref{Sigma(k,tau)}) thereby obtained is transformed into the $\varepsilon_n$ space
in the same way as described for Eqs.\ (\ref{chi(tau)-2})-(\ref{S_nu(L)}).
The fermion Matsubara frequencies $\varepsilon_n\!=\!(2n\!+\!1)\pi T$  ($n\!=\!0,1,2,\cdots$) for $n\!>\! n_{\rm c}\!\sim \! 10$ are chosen so that they are
described approximately by $\varepsilon_n\!\approx\!\sinh x_j$ in terms of equally spaced points $x_j$ ($j\!=\!n_{\rm c}\!+\!2,\cdots,N_\tau$) up to $\varepsilon_n \!\approx\!10^7$; the number of points is $N_\varepsilon\!=\! 50$-$200$.
The self-energy $\Sigma(\vec{k})$ thereby obtained is used to calculate
$\delta G(\vec{k})\!\equiv\!G(\vec{k})\!-\!G_0(\vec{k})$, i.e., the correlation part of Green's function.
This $\delta G(\vec{k})$ is then transformed into the $\tau$ space by the procedure for Eqs.\ (\ref{chi(tau)})-(\ref{I_j,n})
to obtain $\delta G(k,\tau)$ of the next iteration.

Below $T\!=\! T_{1{\rm PR}}$, we need to incorporate the second term of Eq.\ (\ref{dSigma_lambda})
to obtain any finite result. Calculations have been performed for several trial $\lambda_{\rm n}$'s 
to identify the value of $\lambda_{\rm n}$ at which $U\chi_{G\bar{G}}^{}(\vec{0})\!=\!1$ is met.
It has turned out numerically that the value of $U\chi_{G\bar{G}}^{}(\vec{0})$ changes almost linearly with
$\lambda_{\rm n}$ around $U\chi_{G\bar{G}}^{}(\vec{0})\!=\!1$, so that the desired value of  $\lambda_{\rm n}$ 
has been identified with no fundamental difficulties.

Similar calculations have been performed for the superconducting state iteratively
starting from the mean-field solutions.
The convergences are good at low temperatures of $T\!\lesssim\! 0.2T_{{\rm c}0}$
even for the calculations using Eq.\ (\ref{SigmaDelta_lambda-s}).
As we approach $T_{\rm c}\!\approx\!0.0624$, however, the iterations become increasingly unstable.
Hence, we have set $\lambda_{\bar{G}}\!=\!\tilde\lambda_{F}^{}\!>\!0$ and $\lambda_{G}^{}\!=\!\lambda_{F}^{}\!=\!\tilde\lambda_{G}^{}\!=\!0$ from the beginning
on the basis of the results obtained at low temperatures. 
This has enabled us to obtain the convergence  near $T_{\rm c}$, which however has been reached only after 
many iterations.

\end{document}